\documentclass{rmf-d}
\usepackage{nopageno,rmfbib,multicol,times,epsf,amsmath,amssymb,cite}
\usepackage[latin1]{inputenc}
\usepackage[]{caption2}
\usepackage{graphics}

\usepackage{multicol}
\usepackage[demo]{graphicx}

\usepackage{nonfloat}
\newcommand\myfigure[1]{%
\medskip\noindent\begin{minipage}{\columnwidth}
\centering%
#1%
\end{minipage}\medskip}

\def\rmfcornisa{  }
\begin{document}
\title{   Towards precise collider predictions: the Parton Branching method 
\vspace{-6pt}}
\author{ Aleksandra Lelek     }
\address{ Department of Physics, Particle Physics Group, Groenenborgerlaan 171,\\
2020 Antwerp, Belgium   }
\author{ Contribution to the 19th International Conference on Hadron Spectroscopy and Structure (HADRON2021)  }
\address{ }
\author{ }
\address{ }
\author{ }
\address{ }
\author{ }
\address{ }
\author{ }
\address{ }
\maketitle
%\recibido{day month year}{day month year
%\vspace{-12pt}}
\begin{abstract}
\vspace{1em} The collinear factorization theorem, combined with finite-order calculations in perturbative QCD, provides a powerful framework to obtain predictions for many collider observables. However, for observables which  involve multiple energy scales, transverse degrees of freedom cannot be  neglected, and finite-order perturbative calculations have to be combined  with resummed calculations to all orders in the QCD running coupling  in order to obtain reliable theoretical predictions, capable of describing experimental measurements. This is traditionally done either by analytic resummation methods or by parton shower (PS) Monte Carlo (MC) methods. In this talk  we present the Parton Branching (PB) MC method to obtain  QCD collider predictions based on Transverse Momentum Dependent  (TMD) factorization. The PB provides evolution equations for  TMD Parton  Distribution Functions (PDFs) which, upon fitting TMD PDFs to experimental  data, can be used in TMD MC event generators. We present the basic  concepts of the method and illustrate its applications to collider measurements focusing on Drell-Yan (DY) lepton-pair production in different kinematic ranges, from fixed-target to LHC energies. We discuss the latest developments of the method concentrating especially on the matching of next-to-leading-order (NLO) TMD evolution with MC-at-NLO calculations of NLO matrix elements.
   \vspace{1em}
\end{abstract}
\keys{  Transverse Momentum Dependent (TMD) PDFs, resummation, Parton Shower, matching, NLO, Drell-Yan, Drell-Yan + jets, merging  \vspace{-4pt}}

\begin{multicols}{2}
\section{Introduction}
The collinear factorization theorem \cite{Collins:1989gx} is a baseline in obtaining  QCD predictions for production processes at high energy colliders.
Despite its indisputable success, for some observables involving more scales the transverse degrees of freedom of the proton have to be taken into account to obtain precise predictions \cite{Angeles-Martinez:2015sea}.  The
formalism to follow in such scenarios  is the Transverse Momentum Deependent (TMD) factorization theorem which can take the form  of analytical Collins-Soper-Sterman (CSS) approach \cite{Collins:1984kg} or high energy ($k_{\bot}$) factorization \cite{Catani:1990xk,Catani:1990eg}. In practical applications,  the soft gluons resummation is performed by  Parton Shower (PS) algorithms within Monte Carlo (MC) generators.

In recent years a MC method is being developed which makes use of TMD parton distribution functions (PDFs), commonly named as TMDs. The so called Parton Branching (PB) approach  \cite{Hautmann:2017xtx,Hautmann:2017fcj,BermudezMartinez:2018fsv}  turned out to be very flexible and enabled to perform studies in multiple directions \cite{Deak:2018obv,Blanco:2019qbm,VanHaevermaet:2020rro,Hautmann:2019biw,BermudezMartinez:2019anj,BermudezMartinez:2020tys,Jung:2021rak,TaheriMonfared:2021rgc,Keersmaekers:2021cie,Keersmaekers:2021arn,Martinez:2021chk,Abdulhamid:2021xtt}.
 In this work we concentrate on the successful description of Drell-Yan (DY) and DY + jets  data in a wide kinematic range. 

The precision measurements at high energy colliders rely up to a large extend on  
our understanding of the DY processes. 
As a clean production channel, DY is used  in precision electro-weak measurements. DY data are used to extract parton distribution functions (PDFs)  and at low masses and low energies give access to study partons'  intrinsic transverse momentum. 
DY is crucial for our understanding of  QCD evolution and soft gluons resummation. 
DY + jets is of particular interest since the  production channels important for precision measurements and beyond the standard model (BSM) searches 
involve final states with large multiplicities of jets. DY+jets is an important background in many measurements. It is also used in studies on multiparton interactions.

The theoretical description of DY $p_{\bot}$ spectra over wide kinematic regions in energy, mass and $p_{\bot}$ 
requires  a proper matching between the fixed-order perturbative QCD calculations in the high $p_{\bot}$ region (i.e. 
$p_{\bot}
{\raisebox{-.6ex}{\rlap{$\,\sim\,$}} \raisebox{.4ex}{$\,>\,$}}
%\sim 
Q$, where $Q$ is the invariant mass of the DY lepton pair) and soft gluon resummation in the   low $p_{\bot}$ region ($p_{\bot}<<Q$). 

Recent studies \cite{Bacchetta:2019tcu} showed that  the  perturbative fixed-order calculations in collinear factorization are not able to describe the DY $p_{T}$ spectra for $p_{\bot}\sim Q $ at fixed target experiments. In sec.~\ref{sec:DY}  we investigate this issue from the standpoint of  the PB approach. 

The possible impact of TMDs on multi-jet production has just started to be explored. 
In sec.~\ref{sec:tmdhighPt} this topic is addressed, and  
the new PB result for merged DY+jets calculation including higher jets multiplicities is discussed \cite{Martinez:2021chk}. 
\end{multicols}
\myfigure{
\includegraphics[width=0.3\linewidth]{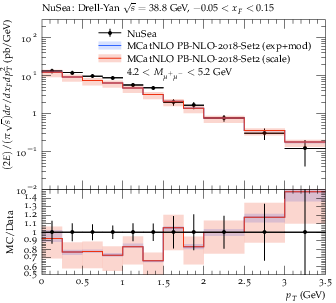}
\includegraphics[width=0.3\linewidth]{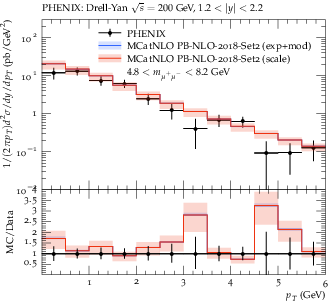}
\includegraphics[width=0.3\linewidth]{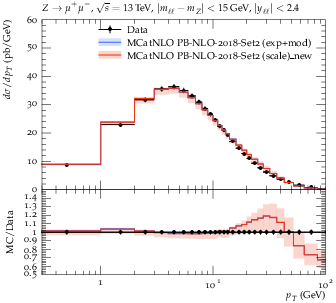}
\figcaption{Predictions for DY $p_{\bot}$ spectra  obtained with MCatNLO+PB TMD compared with data coming from NuSea (left), PHENIX (middle) and CMS (right) experiments  \cite{BermudezMartinez:2020tys}. }
\label{fig:predictions}
}

\myfigure{
\includegraphics[width=0.3\linewidth]{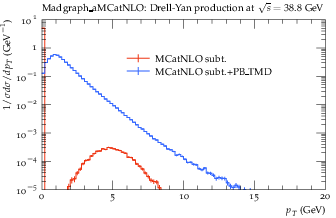}
\includegraphics[width=0.3\linewidth]{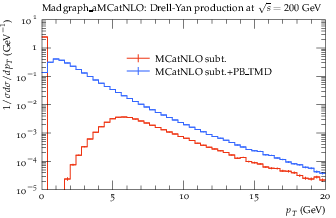}
\includegraphics[width=0.3\linewidth]{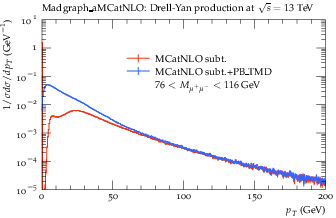}
\figcaption{Subtracted NLO ME from MCatNLO calculation (red) and full MCatNLO+PB TMD calculation (blue) at center of mass energies  corresponding to Fig.~\ref{fig:predictions} \cite{BermudezMartinez:2020tys}. }
\label{fig:subtraction}}
 
\begin{multicols}{2}
\section{The Parton Branching method}
\label{sec:PB}
In the PB method the TMDs are obtained from the PB TMD evolution equation \cite{Hautmann:2017xtx,Hautmann:2017fcj}.
The equation is based on showering \cite{Marchesini:1987cf}  version of the DGLAP equation \cite{Gribov:1972ri,Lipatov:1974qm,Altarelli:1977zs,Dokshitzer:1977sg} where the unitarity picture is used: parton evolution is expressed in terms of  real, resolvable branching probabilities, provided by the real emission DGLAP splitting functions and no-branching probabilities, included via   Sudakov form factors. 
The equation 
describes the change of the TMD $\widetilde{A}_a(x, k_{\bot},\mu)=xA_a(x, k_{\bot},\mu)$ for  all flavors $a$ with the evolution scale $\mu$ calculating  the longitudinal fraction $x$ of the protons' momentum and the transverse momentum $k_{\bot}$ carried by the parton after each branching. 
The starting distribution at initial evolution scale includes the  longitudinal momentum part  and the intrinsic transverse momentum $k_{\bot 0}$.   Using a  parameterization of the HERAPDF2.0 \cite{H1:2015ubc} form for the longitudinal momentum and  a Gaussian in $k_{\bot 0}$,  the parameters of the longitudinal part 
are fitted \cite{BermudezMartinez:2018fsv} to the HERA DIS data using \texttt{xFitter} \cite{Alekhin:2014irh}. The transverse momentum is accumulated at each branching: it is a sum of the intrinsic transverse momentum and all the transverse momenta $q_{\bot}$ emitted in the evolution chain. 
The branchings in the PB method are angular ordered (AO) \cite{Hautmann:2017fcj,Hautmann:2019biw} which allows to perform soft gluons resummation. The AO enters the PB method similarly to \cite{Marchesini:1987cf}, i.e. via 3 elements: 1. relating the DGLAP evolution scale $\mu$ to branching variables $z$ (which is the ratio of the $x$ variables of the partons  propagating towards the hard scattering before and after the branching) and $q_{\bot}$, 2. in the scale of running coupling $\alpha_s(q_{\bot})$ and 3. in the definition of the soft gluons resolution scale $z_M$, which is the maximum value allowed for $z$ variable for a parton to be considered resolvable. 

Delivered PB TMDs  can be  accessed via  TMDlib \cite{Abdulov:2021ivr} and used in TMD MC generators 
(like e.g. CASCADE \cite{Jung:2010si,Baranov:2021uol}). Additionally, 
the PB TMDs can be integrated over $k_{\bot}$ to obtain collinear PDFs $xf(x,\mu)$ (or so called integrated TMDs, iTMDs) which can be then used via LHAPDF \cite{Buckley:2014ana} in collinear physics applications and tools. 
  The  PB parton distributions are applicable in a wide kinematic range of $x$, $k_{\bot}$ and $\mu$.

\section{DY predictions with the PB method }
\label{sec:DY}

The technique to obtain collider predictions with PB TMDs was  proposed in \cite{BermudezMartinez:2018fsv}. In \cite{BermudezMartinez:2019anj} the method was further developed to next-to-leading (NLO) where PB TMDs were combined with NLO matrix element (ME) within the  MADGRAPH5$\_$AMC@NLO (referred later as MCatNLO) \cite{Alwall:2014hca}.

The first step of the generation is performed by MCatNLO. The so called subtracted collinear NLO ME  is generated in the LHE format \cite{Alwall:2006yp} using the integrated PB TMD via LHAPDF. 
In order to avoid possible double counting when combining  NLO ME with PS, the MCatNLO method uses subtraction terms for soft and collinear contributions \cite{Frixione:2002ik}. In the procedure presented here PB TMDs are used instead of PS, their role is however very similar. Because of that the subtraction method has to be used to combine PB TMDs with MCatNLO calculations. The exact form of the  subtraction terms depends on the PS algorithm.  The AO used in PB TMDs is similar to Herwig6 \cite{Corcella:2002jc} so MCatNLO with Herwig6 subtraction is used to combine PB TMDs with MCatNLO. 
In the next step the subtracted collinear ME is supplemented with transverse momentum 
 $k_{\bot}$ by an algorithm in CASCADE, and $k_{\bot}$ is  added to the event record according to the TMD distribution. The TMD used in CASCADE  corresponds to the iTMD from which the ME was initially generated. 
The longitudinal momentum 
fractions of the incoming partons have to be adjusted to conserve energy-momentum and keep the mass of the DY system unchanged. For inclusive observables, like DY $p_{\bot}$ spectrum, 
the whole kinematics is included by using PB TMDs.
\myfigure{\includegraphics[width=.69\columnwidth]{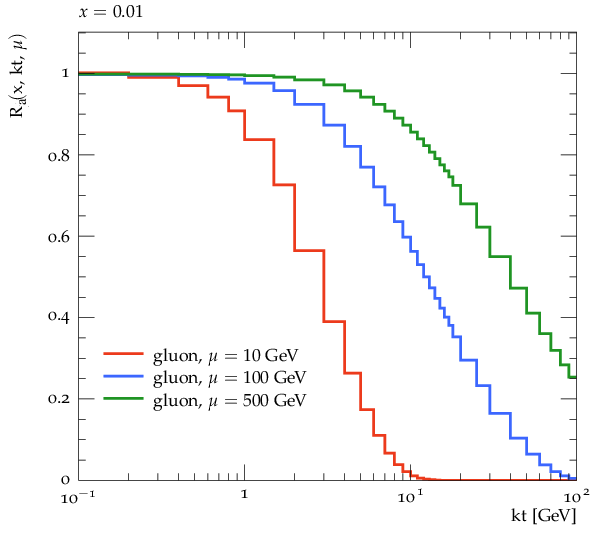}%
\figcaption{The $k_{\bot}$ dependence of the ratio defined in eq.~\ref{eq:ratio} calculated for gluon at $x=0.01$ at different evolution scales $\mu$ \cite{Martinez:2021chk}.}\label{fig:kttail}}

The described procedure using PB-NLO-HERAI+II-2018-set2 TMD PDF~\cite{BermudezMartinez:2018fsv} was applied to obtain predictions for DY $p_{\bot}$ data from different experiments at very different  center of mass energies $\sqrt{s}$ and DY masses \cite{BermudezMartinez:2020tys}:  
NuSea \cite{Webb:2003bj}, R209 \cite{Antreasyan:1981eg}, PHENIX \cite{PHENIX:2018dwt}, ATLAS  \cite{ATLAS:2015iiu}  and CMS \cite{CMS:2019raw}.
The results for  spectra coming from NuSea, PHENIX and CMS are presented in fig.~\ref{fig:predictions} showing a good description  in all these kinematic regimes in small and middle $p_{\bot}$ range. 
To obtain a proper prediction in the high $p_{\bot}$ range, higher jet multiplicities have to be taken into account. This  will be discussed in the next section.

In Fig.~\ref{fig:subtraction} the MCatNLO+PB TMD prediction (blue) is compared to MCatNLO subtracted ME calculation (red) for center of mass energies corresponding to measurements presented in Fig.~\ref{fig:predictions}. At low DY mass and
low energy, the contribution of 
soft gluon emissions contained in PB TMDs is crucial to describe the
data  even in the region of $p_{\bot} \sim Q$. The situation is different at LHC energies and larger masses,  
where  the contribution from soft
gluons in the region of $p_{\bot} \sim Q$ is small and the spectrum is governed by hard real emission. This confirms the observation from the literature that   perturbative fixed-order calculations in collinear factorization are not able to describe DY $p_{\bot}$ spectra at fixed target experiments in the region of $p_{\bot}\sim Q $.

\section{TMD effects at high $p_{\bot}$}
\label{sec:tmdhighPt}
It is commonly known that TMD effects play a role at scales of order of few GeV. The question remains if they can be also important at higher scales. 
In the PB approach the TMD at the initial evolution scale $\mu\sim \mathcal{O}(1 \;\rm{GeV})$   is a Gaussian with width $\sigma$,  $\Lambda_{QCD}<\sigma < \mathcal{O}(1\; \rm{GeV})$. During the evolution the transverse momentum $k_{\bot}$ is  accumulated  in each step which leads to TMD broadening.
In the PB method, the PDFs can be obtained from a TMD by integration over the transverse momentum $\widetilde{f}_{a}(x,\mu^2) = \int \textrm{d}k_{\bot}^2\widetilde{A}_a(x, k_{\bot}, \mu^2)$. In order to estimate what is the probability of a parton $a$ with momentum fraction $x$ to acquire the transverse momentum higher than $k_{\bot}$  at a given evolution scale $\mu$ one can define a ratio \cite{Martinez:2021chk} 
\begin{equation}
\label{eq:ratio}
 R_a (x, k_{\bot}, \mu^2) = \frac{\int_{k_{\bot}^2}^{\infty} \textrm{d}k_{\bot}^{\prime 2}\widetilde{A}_a(x, k_{\bot}^{\prime}, \mu^2)}{\int \textrm{d}k_{\bot}^{\prime 2}\widetilde{A}_a(x, k_{\bot}^{\prime}, \mu^2)}
\end{equation}
which is shown in fig.~\ref{fig:kttail} for 3 different evolution scales for gluon at $x=0.01$. 
From this figure one can estimate e.g. that at $\mu=100\;\rm{GeV}$ the probability of a gluon to have a transverse momentum higher than  $20\;\rm{GeV}$ is $30\%$. In other words, at LHC the contribution from the $k_{\bot}$-broadening of the TMD
to e.g. the emission of an extra jet in a process characterized by a hard scale $\mu$ being the transverse momentum of a jet  is comparable to emissions from hard matrix element \cite{Martinez:2021chk}. 
\myfigure{\includegraphics[width=.79\columnwidth]{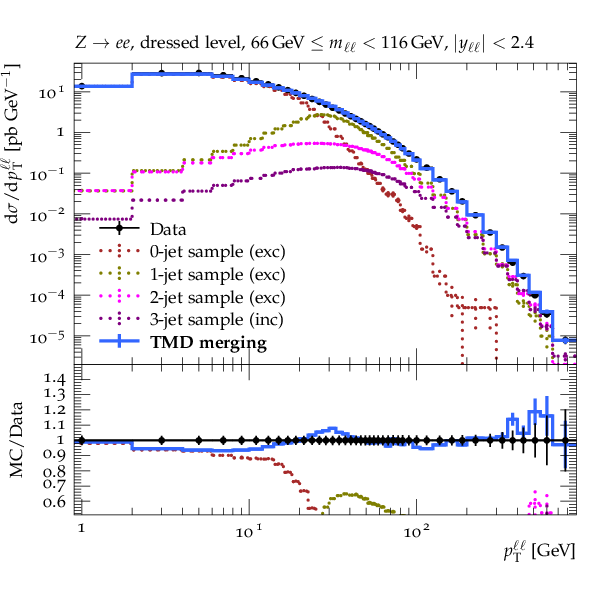}%
\figcaption{The fully
TMD-merged calculation, as well as separate contributions
from the different jet samples compared to $8\;\rm{TeV}$ATLAS data for DY $p_{\bot}$ spectrum \cite{Martinez:2021chk}.}\label{fig:MLMresults}} 
It was shown in \cite{BermudezMartinez:2019anj} that the proper description of the high $p_{\bot}$ part of the DY spectrum requires the contribution from higher jet multiplicities.
In \cite{Martinez:2021chk} the standard MLM merging procedure \cite{Mangano:2006rw,Alwall:2007fs} was extended to the TMD case at leading order (LO). The TMD merging was applied to Z boson production in association with jets. 
Predictions coming from contributions from different jet samples and fully merged calculation 
 is compared to $8\;\rm{TeV}$ ATLAS data \cite{ATLAS:2015iiu}
in fig.~\ref{fig:MLMresults}. The merged calculation gives a good description of the data in the whole $p_{\bot}$ range of the DY spectrum. One can see that at low $p_{\bot}$ Z+0 jet sample is the main contribution to the spectrum and the importance of the higher multiplicities increases with $p_{\bot}$. 
In fig.~\ref{fig:MLMresultsmultiplicities} the predictions for jet multiplicity in Z+jets production obtained with merged calculation as well separate contributions form each jet multiplicity sample is compared with $13\;\rm{TeV}$ ATLAS data \cite{ATLAS:2017sag}. The agreement with data is excelent, also for the jet multiplicities higher than the maximum number of jets generated at the ME level.

\myfigure{\includegraphics[width=.79\columnwidth]{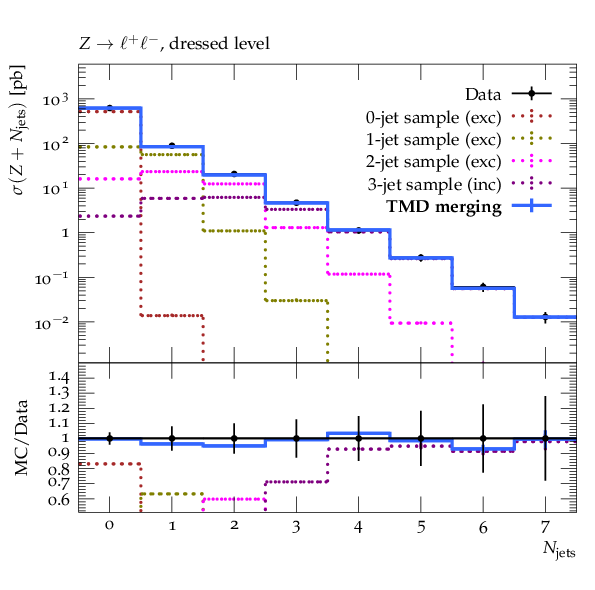}%
\figcaption{The fully
TMD-merged calculation, as well as separate contributions
from the different jet samples compared to $13\;\rm{TeV}$ATLAS data  for jet multiplicity in Z+jets production \cite{Martinez:2021chk}.}\label{fig:MLMresultsmultiplicities}}

\section{Conclusions}

 A central part of the LHC physics program are the precision strong and electro-weak measurements. For these, 
 the accuracy of  theoretical predictions of DY and DY+jets data is very important.

In this work the predictions obtained from the PB method  for  the DY $p_{\bot}$ measurements over a wide range in energy, DY mass and $p_{\bot}$ are shown.
At low mass and low energy  both fixed-order QCD and all-order soft gluon emissions are important to describe the DY $p_{\bot}$, and the reliability 
of the theoretical predictions depends on matching between those two elements.
In the presented method in low and middle $p_{\bot}$ range the NLO ME  was combined with PB TMDs with the matching performed according to MCatNLO method.
This allowed one to confirm the observation  from the literature that  perturbative fixed-order calculations in collinear factorization are not able to describe DY $p_{\bot}$ spectra at fixed target experiments in the region of $p_{\bot}\sim Q $. Moreover it was noticed that the contribution from soft gluons included in PB TMDs is essential to describe these data. 
The situation changes with the center of mass energy: at LHC  the region of $p_{\bot}\sim Q$ is well described by the   collinear NLO calculation. 

It was shown that because of the TMD broadening, the TMD effects cannot be neglected  at high $p_{\bot}$. The high $p_{\bot}$ region of the DY spectrum requires taking into account contributions from higher orders. The new TMD merging method was recently developed at LO to merge different jet multiplicities. With this method a very good description of the LHC DY $p_{\bot}$ spectrum in the whole $p_{\bot}$ region was obtained. 
The study opens the possibility to
further investigate  TMD effects at
the level of exclusive jet observables and in the region of
the highest $p_{\bot}$, where e.g.
signals of BSM physics could be largest.

\section*{Acknowledgments}
I acknowledge funding by Research Foundation-Flanders (FWO) (application number: 1272421N).

\end{multicols}

\medline
\begin{multicols}{2}

\end{multicols}

\end{document}